\documentclass[preprintnumbers,
amsmath,
prd,nofootinbib,floatfix,12pt,
]{revtex4}

\newcommand\beq{\begin{eqnarray}}
\newcommand\eeq{\end{eqnarray}}
\def\Vresummed{\widehat{V}}
\def\gptwo{g^{\prime 2}}
\def\gpfour{g^{\prime 4}}

\def\lnbar{\overline{\ln}}
\def\lsim{\mathrel{\rlap{\lower4pt\hbox{$\sim$}}
    \raise1pt\hbox{$<$}}}                
\def\gsim{\mathrel{\rlap{\lower4pt\hbox{$\sim$}}
    \raise1pt\hbox{$>$}}}            

\allowdisplaybreaks
\def\MSbar{\overline{\rm MS}}

\usepackage{graphicx}
\usepackage{setspace}
\begin{document}
\renewcommand{\theequation}{\arabic{section}.\arabic{equation}}
\renewcommand{\thefigure}{\arabic{section}.\arabic{figure}}
\renewcommand{\thetable}{\arabic{section}.\arabic{table}}

\title{\large \baselineskip=16pt 
Taming the Goldstone contributions to the 
effective potential}

\author{Stephen P.~Martin}
\affiliation{
\mbox{\it Department of Physics, Northern Illinois University, DeKalb IL 60115}, and\\
\mbox{\it Fermi National Accelerator Laboratory, P.O. Box 500, Batavia IL 60510}
}

\begin{abstract}\normalsize \baselineskip=15pt
The standard perturbative effective 
potential suffers from two 
related problems of principle involving the field-dependent Goldstone 
boson squared mass, $G$. First, in general $G$ can be negative, and it 
actually is negative in the Standard Model; this leads to imaginary 
contributions to the effective potential that are not associated with a 
physical instability, and therefore spurious. Second, in the limit that 
$G$ approaches zero, the effective potential minimization condition is 
logarithmically divergent already at two-loop order, and has 
increasingly severe power-law singularities at higher loop orders. I 
resolve both issues by resumming the Goldstone boson 
contributions to the effective potential. 
For the resulting resummed effective potential, the minimum value and
the minimization condition that gives the vacuum expectation value
are obtained in forms that do not involve $G$ at all.
\end{abstract}

\maketitle
\tableofcontents
\baselineskip=18pt
\setcounter{footnote}{1}
\setcounter{figure}{0}
\setcounter{table}{0}

\newpage

\section{Introduction\label{sec:intro}}
\setcounter{equation}{0}
\setcounter{figure}{0}
\setcounter{table}{0}
\setcounter{footnote}{1}

A resonance with mass about 126 GeV
and properties expected of a minimal Standard Model Higgs scalar boson 
has been discovered 
\cite{ATLASHiggs,CMSHiggs,ATLAScombination,CMScombination}
at the Large Hadron Collider. 
One of the theoretical tools useful for understanding the electroweak
symmetry breaking dynamics of the minimal Standard Model and its extensions
is the effective potential
\cite{Coleman:1973jx,Jackiw:1974cv,Sher:1988mj},
which can be used to relate the Higgs field vacuum expectation value (VEV) to
the fundamental Lagrangian parameters, and to observable quantities such 
as the masses of the Higgs bosons, the top quark, and the 
masses and interactions of the $W$ and $Z$ bosons.
On general grounds, the Standard Model Lagrangian parameter should be
obtained as accurately as possible.
It may be possible to discern the difference between the minimal Higgs
Standard Model and more complicated theories, and to gain hints about
the mass scale of new physics, and the stability of the Standard Model vacuum state. 
An interesting feature of the Higgs mass 
is that the potential is close to a metastable region
associated with a very small Higgs self-interaction 
coupling at very large energy scales.
Some studies of the stability condition 
that were made before the Higgs discovery are 
refs.~\cite{Sher:1988mj,Lindner:1988ww,Arnold:1991cv,Ford:1992mv,Casas:1994qy,Espinosa:1995se,Casas:1996aq,Isidori:2001bm,Espinosa:2007qp,ArkaniHamed:2008ym,Bezrukov:2009db,Ellis:2009tp}, 
and some of the analyses following the Higgs discovery are 
refs.~\cite{EliasMiro:2011aa,Alekhin:2012py,Bezrukov:2012sa,Degrassi:2012ry,Buttazzo:2013uya}.

To fix notation, write the complex doublet Higgs field as
\beq
\Phi(x) = \begin{pmatrix}
\frac{1}{\sqrt{2}} [\phi + H(x) + i G^0(x)]
\\
G^+(x)
\end{pmatrix}.
\eeq
Here $\phi$ is the real background field, about which are expanded 
the real Higgs quantum field $H$, 
and the real neutral and complex charged Goldstone boson fields
$G^0$ and $G^+ = G^{-*}$.
The Lagrangian for the Higgs kinetic term and its 
self-interactions are given by
\beq
{\cal L} =  -\partial^\mu \Phi^\dagger \partial_\mu \Phi 
-\Lambda -m^2 \Phi^\dagger \Phi -\lambda (\Phi^\dagger \Phi)^2 ,
\eeq
where $m^2$ is the Higgs squared mass parameter, 
and $\lambda$ is the Higgs self-coupling 
in the normalization to be used in this paper, and the 
metric signature is ($-$$+$$+$$+$). 
The field-independent vacuum energy density $\Lambda$ 
is necessary for renormalization scale invariance of the effective 
potential and a proper treatment of renormalization group 
improvement 
\cite{Yamagishi:1981qq,Einhorn:1982pp,Kastening:1991gv,Bando:1992np,Ford:1992mv,Einhorn:2007rv}. 
The Lagrangian also includes a top-quark Yukawa coupling $y_t$, and 
$SU(3)_c \times SU(2)_L \times U(1)_Y$ gauge couplings 
$g_3$, $g$, and $g'$. The other Yukawa couplings are very small, and can make 
only a very minor difference even at 1-loop order, and so are neglected.
All of the Lagrangian parameters as well as the background field $\phi$
depend on the
$\MSbar$ renormalization scale $Q$, and 
logarithms of dimensional quantities are written below in terms of
\beq
\lnbar(x) &\equiv& \ln(x/Q^2).
\eeq

The effective potential can be evaluated in perturbation theory and written as:
\beq
V_{\rm eff}(\phi) &=& \sum_{\ell= 0}^\infty \frac{1}{(16\pi^2)^\ell} 
V_{\ell}
.
\eeq
In this paper, the power of $1/16 \pi^2$ is used as a signifier for 
the loop order $\ell$.
The tree-level potential is given by: 
\beq
V_0
&=& \Lambda + \frac{m^2}{2} \phi^2 + \frac{\lambda}{4}\phi^4.
\label{eq:Vzero}
\eeq
The radiative corrections for $\ell \geq 1$ are obtained from the sum of 1-particle-irreducible vacuum graphs. The Landau gauge is most often used for effective potential calculations, because of the 
simplifications that the gauge-fixing parameter 
is not renormalized and there is
no mixing between the longitudinal vector modes and the Goldstone modes. 
In Landau gauge and the $\MSbar$ renormalization scheme based on dimensional
regularization, the 1-loop order contribution to the 
Standard Model effective potential is:
\beq
V_{1}
&=& 3 f(G) + f(H) -12 f(t)  + 6 f(W) + W^2 + 3 f(Z) + \frac{1}{2} Z^2,
\label{eq:Vone}
\eeq
where 
\beq
f(x) &\equiv& \frac{x^2}{4} \left [\lnbar(x) - 3/2 \right ],
\eeq
and the field-dependent running squared masses are 
\beq
G &=& m_{G^0}^2 = m_{G^\pm}^2 = m^2 + \lambda \phi^2 ,
\\
H &=& m_H^2 = m^2 + 3 \lambda \phi^2,
\label{eq:defH}
\\
t&=& m_t^2 = y_t^2 \phi^2/2,
\\
W&=& m_W^2 = g^2 \phi^2/4,
\\
Z&=&m_Z^2 = (g^2 + g^{\prime 2}) \phi^2/4.
\eeq
The contributions $W^2 + \frac{1}{2}Z^2$ in eq.~(\ref{eq:Vone}) are due 
to the fact that in dimensional regularization the vector fields have $4 
- 2 \epsilon$ components rather than 4.

The full two-loop order contribution $V_{2}$ 
in Landau gauge was worked out by Ford, Jack and Jones
in \cite{Ford:1992pn} for the Standard Model, and for more general
theories (including softly-broken supersymmetric models, where 
regularization by dimensional
reduction is used instead of dimensional regularization) in 
\cite{Martin:2001vx}.
The 3-loop contribution $V_{3}$ for the Standard Model 
was obtained in \cite{Martin:2013gka} in the approximation
that the strong and top Yukawa couplings are much larger than the electroweak
couplings and other Yukawa couplings. For completeness, these results are 
compiled in an Appendix of the present paper in a notation compatible with the
discussion below.

Results for more general gauge-fixing conditions are apparently 
only available at 1-loop order
at present. The effective potential itself is gauge-fixing dependent, but 
physical observables derived from it are not. For discussions of the
gauge-fixing dependence of the effective potential from various points of view
see 
refs.~\cite{Dolan:1974gu,Kang:1974yj,Fischler:1974ue,Frere:1974ia,
Nielsen:1975fs,Fukuda:1975di,Aitchison:1983ns,Kobes:1990dc,Metaxas:1995ab,Gambino:1999ai,Alexander:2008hd,
Loinaz:1997td,DelCima:1999gg,Patel:2011th,DiLuzio:2014bua,Nielsen:2014spa}.

The purpose of this paper is to resolve two issues of principle 
regarding contributions to the effective potential involving the 
Goldstone bosons. Note that the condition $G=0$ marks the minimum of the 
tree-level potential 
$V_{0}$ 
in eq.~(\ref{eq:Vzero}). However, in 
general $G$ will be non-zero at the minimum of the full effective 
potential.

\begin{figure}[t]
\begin{minipage}[]{0.61\linewidth}
\includegraphics[width=0.92\linewidth,angle=0]{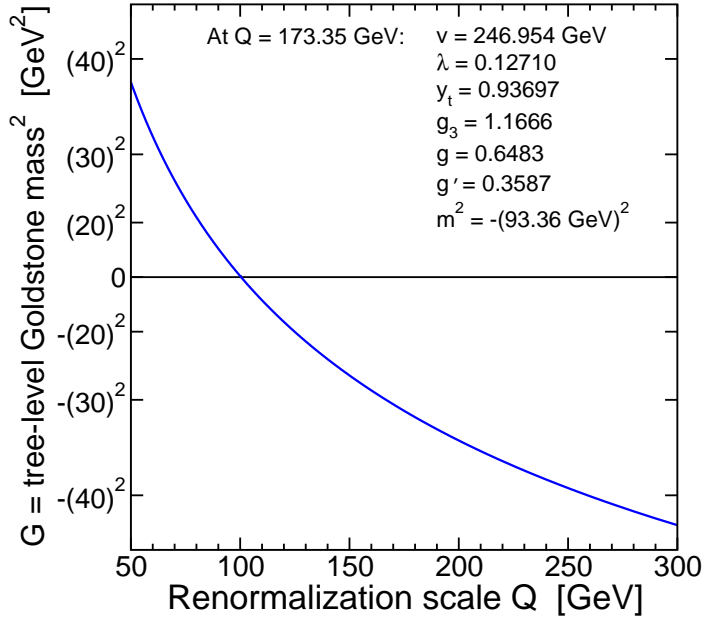}
\end{minipage}
\begin{minipage}[]{0.38\linewidth}
\caption{\label{fig:Grun}
The running of the Landau gauge 
Standard Model Goldstone boson squared mass $G$, 
evaluated at the minimum of the effective potential, 
as a function of the renormalization scale $Q$, 
for the choice of input parameters specified in the text.}
\end{minipage}
\end{figure}

The first problem of principle is that there is no reason why $G$ cannot 
be negative at the minimum of $V_{\rm eff}$, depending on the choice of 
renormalization scale $Q$. Indeed, in the case of the Standard Model, 
$G$ is negative for the perfectly reasonable range $Q \gsim 100$ GeV. 
This is illustrated in Figure 
\ref{fig:Grun} for a typical numerical choice of the parameters (in this 
case taken from ref.~\cite{Buttazzo:2013uya}):
\beq
\lambda(M_t) &=& 0.12710,
\label{eq:inputlambda}
\\
y_t(M_t) &=& 0.93697,
\\
g_3(M_t) &=& 1.1666,
\\
m^2(M_t) &=& -\mbox{(93.36 GeV})^2,  
\\
g(M_t) &=& 0.6483,
\\
g'(M_t) &=& 0.3587.
\label{eq:inputgp}
\eeq
where $Q = M_t = 173.35$ GeV is the input scale.
The problem is that the $\lnbar(G)$ terms give rise to an imaginary part 
of $V_{\rm eff}$. In general, a complex value of $V_{\rm eff}$ 
at its minimum reflects an 
instability \cite{Weinberg:1987vp}, 
but there is no physical instability here. In 
practice, this unphysical imaginary part of $V_{\rm eff}$ has simply been 
ignored, and the real part is minimized. 
In principle, this imaginary part is spurious, 
and a way of making this plain is desirable.

The second problem of principle is that when calculated to fixed loop order 
$\ell$, the effective potential diverges for $G \rightarrow 0$. For $\ell = 
1,2,3$, one finds \cite{Martin:2013gka} 
that 
$
V_{\ell}
\sim G^{3-\ell}\, \lnbar G$, while for
$\ell \geq 4$ one has 
$
V_{\ell}
\sim G^{3 - \ell}$. (The precise results
for the leading behavior as $G \rightarrow 0$ 
will be obtained in the next section.)
Thus the effective potential itself has a logarithmic singularity
at 3-loop order, with increasingly severe power-law singularities for higher 
loop orders. The first derivative of the effective potential, which is
used in the minimization condition that governs the vacuum expectation value of 
the Higgs field, therefore 
has a logarithmic singularity at 2-loop order, and power law 
singularities for higher loop orders. As we will see below, the numerical 
impact of these singularities at the known loop order, $\ell=3$, is quite small
unless one carefully tunes $G\approx 0$ by the choice of renormalization scale. 
However, this is still disturbing as a matter of principle. Indeed, one might 
have expected that a choice of $Q$ that makes $G$ very small would be a 
particularly good choice, because the pole mass of the Landau gauge
Goldstone bosons is 0,
and so $G=0$ corresponds to choosing a renormalization scale such that 
the radiative corrections to it vanish. However, if one uses the 
usual perturbative effective potential truncated at any loop order beyond
1-loop order, 
this is the one renormalization 
scale choice that one must {\em not} make.

The purpose of this paper is to show how the above two problems of 
principle are eliminated by doing a resummation to all loop orders of the 
Goldstone contributions that are leading as $G \rightarrow 0$.

\section{Effective potential contributions for small $G$
\label{sec:renormalized}}
\setcounter{equation}{0}
\setcounter{figure}{0}
\setcounter{table}{0}
\setcounter{footnote}{1}

\begin{figure}[t]
\begin{center}
\includegraphics[width=\linewidth,angle=0]{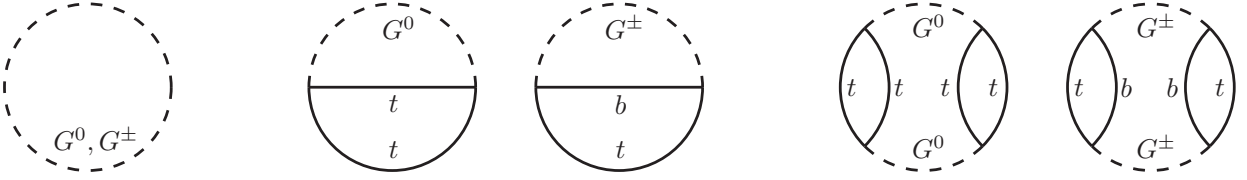}
\end{center}
\vspace{-0.5cm}
\caption{\label{fig:G123}
These Feynman diagrams give the leading non-zero contribution to $V_{\rm eff}$
as $G \rightarrow 0$, at the leading order in $y_t$, for 1-loop, 2-loop, and 3-loop orders.} 
\end{figure}

\begin{figure}[t]
\begin{center}
\includegraphics[width=0.92\linewidth,angle=0]{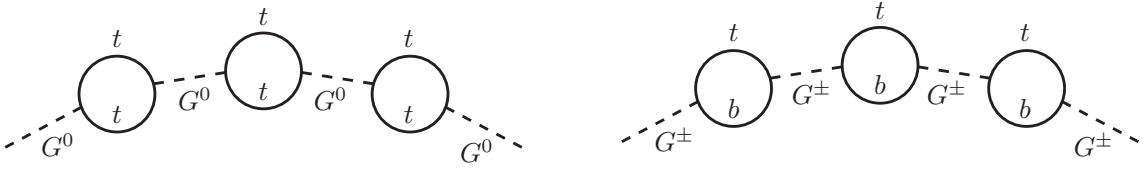}
\end{center}
\vspace{-0.7cm}
\caption{\label{fig:daisychains}
Chains of Goldstone boson propagators interspersed with top and top/bottom loops. Rings of these 
(and similar diagrams involving loops with $Z$, $W$, and $H$) give rise to the most singular contributions as $G \rightarrow 0$, at any given loop order.} 
\end{figure}

For the known contributions to the effective potential, the leading behavior
as $G \rightarrow 0$ can be isolated as follows. First, at one loop order, we have immediately from eq.~(\ref{eq:Vone}) that the Goldstone 
boson contributions are:
\beq
V_{1} 
&=& 3 f(G) + \ldots = \frac{3}{4} G^2 [\lnbar(G) - 3/2] + \ldots
\> .
\label{eq:V1G}
\eeq
In eq.~(\ref{eq:V1G}), the ellipses represent terms with no $G$ dependence.
At 2-loop order, using the expansions of the 2-loop integral
function $I(x,y,z)$ 
for small $G$ given in eqs.~(\ref{eq:expI00G})-(\ref{eq:expIGxx}) of the 
Appendix 
(from eqs.~(2.29)-(2.31) of ref.~\cite{Martin:2001vx}), 
one finds that
\beq
V_{2}
&=& \frac{3}{2} \Delta_1 
G \lnbar(G)
+ \ldots
,
\label{eq:V2G}
\eeq
where
\beq
\Delta_1 &=&  -6 y_t^2 A(t) + 3 \lambda A(H) 
+ \frac{g^2}{2} [3 A(W) + 2 W] +
\frac{g^2 + g^{\prime 2}}{4} [3 A(Z) + 2 Z],
\label{eq:Delta1}
\eeq
and the 1-loop integral function is defined by
\beq
A(x) &\equiv& x (\lnbar x - 1).
\label{eq:defA}
\eeq
(In ref.~\cite{Martin:2001vx}, $A(x)$ was called $J(x)$.)  
In eq.~(\ref{eq:V2G}), the ellipses represent terms independent of $G$,
terms of order $G^2$,
as well as those proportional to $G$ with no $\lnbar(G)$.
Reading directly from eq.~(4.38) of ref.~\cite{Martin:2013gka}, one obtains
the leading behavior as $G \rightarrow 0$ at 3-loop order:
\beq
V_{3}
&=& 27 y_t^4\, A(t)^2 \,\lnbar(G)  + \ldots
\>.
\label{eq:V3G}
\eeq
As noted in ref.~\cite{Martin:2013gka}, the contributions 
above (that involve $y_t$, in the 2-loop case)
come from the diagrams shown in Figure \ref{fig:G123}.
At higher loop orders, the leading contribution as $G \rightarrow 0$ 
and at leading order in $y_t$ comes from
$\ell$-loop order vacuum diagrams consisting of a ring of $\ell-1$
Goldstone boson propagators, interspersed with $\ell-1$ top (for $G^0$) 
or top/bottom (for $G^\pm$) 1-loop subdiagrams, 
as shown in Figure \ref{fig:daisychains}.

More generally still, the 1-loop subdiagrams can be 
any 1-particle irreducible sub-diagrams that 
involve a squared mass scale that can be treated as parametrically large 
compared to $G$. In the Standard Model, these includes 1-loop subdiagrams
containing $Z$, $W$, and $H$ bosons, as well as multi-loop subdiagrams.

To obtain the leading behavior as $G \rightarrow 0$, one considers the 
contributions from these diagrams from integrating 
momenta $p^\mu$ flowing around the large rings 
with $p^2$ small compared to the squared mass scale set by the
1-particle-irreducible  sub-diagrams. 
Then the 1-particle-irreducible sub-diagram contributions can be treated
as just constant squared-mass insertions in the Goldstone boson 
propagators, reducing the 
calculation to a 1-loop integration. The sum of the
resulting leading $G \rightarrow 0$
contributions to 
$V_{\rm eff}$ at each loop order, including the $\MSbar$ counterterms, is:
\beq
V_{\rm eff} &=& \frac{3}{16 \pi^2} \sum_{n=0}^\infty 
\frac{\Delta^n}{n!}
\left (\frac{d}{d G}\right )^{n}
f(G) + \ldots ,
\label{eq:Veffallorders}
\eeq
where $n = \ell-1$ where $\ell$ is the loop order, and
\beq
\Delta &=& \frac{1}{16\pi^2} \Delta_1 + \frac{1}{(16\pi^2)^2} \Delta_2 
+ \frac{1}{(16\pi^2)^3} \Delta_3
+ \ldots\>.
\eeq
Equation (\ref{eq:Veffallorders}) generalizes eqs.~(\ref{eq:V1G}),
(\ref{eq:V2G}), and (\ref{eq:V3G}). For the purposes of 
making this particular comparison, $\Delta_2$
can be dropped, because ref.~\cite{Martin:2013gka} retained only the leading
order in $y_t$ at 3 loops. For the same reason, all but the $y_t^2$ term in
$\Delta_1$ can be dropped in comparing the 3-loop order contributions.
However, in the future, if more terms are calculated in $V_{\rm eff}$ at 
3-loop order and beyond, then those contributions would become pertinent,
as would contributions from other diagrams.

The origin of the prefactor 3 in eq.~(\ref{eq:Veffallorders}) 
is a factor of 2 for 
the $G^\pm$ rings, and a factor 
1 for the $G^0$ ring. Despite the fact that the 1-particle-irreducible
subdiagrams are different for these two classes of diagrams 
(e.g.~involving top/bottom loops for $G^\pm$, and 
top loops for $G^0$), the quantity $\Delta$
is the same in both cases.

Note that $f'(G) = \frac{1}{2}A(G) = \frac{1}{2} G [\lnbar(G)-1]$, 
and $f''(G) = \frac{1}{2}\lnbar(G)$,
and the $n$th derivative is $f^{(n)}(G) = 
\frac{1}{2} (-1)^{n-1}\, (n-3)!\, G^{2-n}$ for $n\geq 3$.
Therefore, the leading singular behavior as $G \rightarrow 0$ is
\beq
V_3 &=& \frac{3}{4} (\Delta_1)^2 \,\lnbar(G) ,
\\
V_\ell &=& 
-\frac{3\, (-\Delta_1)^{\ell-1}}{2 (\ell-1)(\ell-2)(\ell-3)\, G^{\ell-3}}
\qquad\quad \mbox{(for $\ell > 3$).}
\eeq

\section{Resummation of leading Goldstone contributions\label{sec:resum}}
\setcounter{equation}{0}
\setcounter{figure}{0}
\setcounter{table}{0}
\setcounter{footnote}{1}

The contributions to $V_{\rm eff}$ in eq.~(\ref{eq:Veffallorders}) 
from all loop orders $\ell = n+1$ resum to
\beq
V_{\rm eff}&=& \frac{3}{16 \pi^2} f(G + \Delta)  + \ldots \> .
\label{eq:Veffsum}
\eeq
Thus, the result of summing all orders in perturbation theory 
yields a result which is well-behaved for all $G$, unlike 
the result obtained if it is truncated at any finite order 
in perturbation theory. In fact, at the minimum of the full effective
potential,
$G + \Delta = 0$, and the result of the resummation of this class of terms
vanishes. Therefore, 
if $V_{\rm eff}$ has been evaluated at some finite $\ell$-loop 
order in perturbation theory, a sensible result can be obtained 
by simply subtracting off the first $\ell$ terms 
in the series eq.~(\ref{eq:Veffallorders}), and then adding
back in the resummed version of the same series, eq.~(\ref{eq:Veffsum}).
If the effective potential $V_{\rm eff}$ has been calculated to loop order 
$\ell$, then the resummed effective potential is
\beq
\Vresummed_{\rm eff} &=& V_{\rm eff} + \frac{3}{16 \pi^2} \left [
f(G + \Delta) - \sum_{n=0}^{\ell-1} 
\frac{\Delta^n}{n!}
\left (\frac{d}{d G}\right )^{n}
f(G) \right ] .
\label{eq:genresum}
\eeq
The result is free of 
the offending leading singular contributions 
as $G \rightarrow 0$. 

Further resummation may be necessary to account 
for other terms that are singular as $G \rightarrow 0$ but sub-leading at
a given loop order, 
but these do not arise in the
effective potential approximation that has been calculated so far.
If we use $V_{\rm eff}$ to refer to the 
usual full 2-loop and leading 3-loop Standard Model effective 
potential as computed in
refs.~\cite{Ford:1992pn} and \cite{Martin:2013gka}, 
then the appropriate resummed version from eq.~(\ref{eq:genresum}) is:
\beq
\Vresummed_{\rm eff} &=& V_{\rm eff} + 
\frac{3}{16 \pi^2} \left [f(G + \Delta) - f(G) \right ]
-\frac{1}{(16 \pi^2)^2} \frac{3\Delta_1 }{2} A(G) 
-\frac{1}{(16 \pi^2)^3} 27 y_t^4 \,A(t)^2\, \lnbar(G) .
\nonumber \\ &&
\label{eq:Veffresummed}
\eeq
Here I have taken $(\Delta_1)^2 = 36 y_t^4 A(t)^2$ in the 3-loop 
part, because only the leading order in $y_t$ for $V_{\rm eff}$ 
was included in 
ref.~\cite{Martin:2013gka}. For the same reason, the 3-loop order term 
involving $\Delta_2$ is dropped here.
The effect of the 1-loop correction in eq.~(\ref{eq:Veffresummed}) 
is to replace the tree-level field-dependent
Goldstone boson squared mass by
its pole squared mass, which vanishes at the minimum of the 
full effective potential in Landau gauge. 
The 3-loop order term simply cancels the 
corresponding $\lnbar(G)$ contribution found in ref.~\cite{Martin:2013gka}.
I propose that the resummed version of the effective potential,
$\Vresummed_{\rm eff}$, should be used instead of the usual $V_{\rm eff}$.

\section{Minimization condition for the effective potential\label{sec:min}}
\setcounter{equation}{0}
\setcounter{figure}{0}
\setcounter{table}{0}
\setcounter{footnote}{1}

For the usual effective potential $V_{\rm eff}$, the minimization
condition that relates the vacuum expectation value 
$v = \phi_{\rm min}$ to the Lagrangian parameters is
\beq
G \>=\> m^2 + \lambda v^2 
&=&
-\frac{1}{16 \pi^2} \delta_1 
-\frac{1}{(16 \pi^2)^2} \delta_2 
-\frac{1}{(16 \pi^2)^3} \delta_3
-\ldots ,
\eeq
where the correction at $\ell$-loop order is:
\beq
\delta_\ell &=&
\frac{1}{\phi}\frac{\partial}{\partial \phi} 
V_{\ell}
\Bigl |_{\phi = v} .
\eeq
(In the remainder of this section, I will use $\phi=v$, because
all equations hold only at the minimum of the potential.)
Explicitly, at 1-loop order, one has from eq.~(\ref{eq:Vone}):
\beq
\delta_1 &=& 3 \lambda A(G) -6 y_t^2 A(t) + 
3 \lambda A(H)  + \frac{g^2}{2} [3A(W) +2W]
+ \frac{g^2 + g^{\prime 2}}{4} [3A(Z) +2Z],
\phantom{xxx}
\label{eq:delta1}
\eeq
and the higher loop order contributions can be obtained by
taking derivatives of the
results of ref.~\cite{Ford:1992pn}
at 2-loop order, and
from ref.~\cite{Martin:2013gka} at 3-loop order for terms that 
are leading order in $g_3$ and $y_t$.
Note that $\delta_1$ differs from the quantity $\Delta_1$, 
given in eq.~(\ref{eq:Delta1}) above, only by the inclusion of the
first term, $3 \lambda A(G)$. 
At higher loop orders $\ell \geq 2$, it is useful to
note the leading dependence on $G$ as $G \rightarrow 0$:
\beq
\delta_2 &=& 3\lambda \Delta_1 \lnbar(G) - 9 y_t^4 \lnbar(t) A(G) + \ldots
,
\label{eq:delta2}
\\
\delta_3 &=& 54 y_t^4 \left [
\frac{\lambda A(t)^2}{G} 
+ y_t^2 A(t) \lnbar(t) \lnbar(G)
\right ] + \ldots \>.
\label{eq:delta3}
\eeq
Equation (\ref{eq:delta2}) can be obtained using 
eqs.~(\ref{eq:V2decomp})-(\ref{eq:fgauge}) and
(\ref{eq:expI00G})-(\ref{eq:expIGxx}) in the Appendix,
and eq.~(\ref{eq:delta3}) can be obtained from eq.~(\ref{eq:V3rensim}).
In $\delta_2$, the ellipses includes 
terms that do not depend on $G$, 
terms with a linear factor of $G$ but suppressed by $g, g', \lambda$
or not containing $\lnbar(G)$,  and terms quadratic or
higher order in $G$. In $\delta_3$, the ellipses represents terms that
are finite as $G \rightarrow 0$ or suppressed by $g, g', \lambda$.

The condition for minimization of $\Vresummed_{\rm eff}$ 
defined in eq.~(\ref{eq:Veffresummed}) is:
\beq
G \>=\> m^2 + \lambda v^2 &=& 
-\frac{1}{16 \pi^2} \Delta_1
-\frac{1}{(16 \pi^2)^2} \Delta_2 
-\frac{1}{(16 \pi^2)^3} \Delta_3
- \ldots \> ,
\eeq
where
\beq
\Delta_\ell &=&
\frac{1}{v}\frac{\partial}{\partial v} 
\Vresummed_{\ell}
,
\eeq
with $\displaystyle
\Vresummed_{\rm eff} 
= \sum_{\ell} \frac{1}{(16 \pi^2)^\ell}
\Vresummed_{\ell}
$. 
Consider the difference between the minimization conditions for
$V_{\rm eff}$ and $\widehat V_{\rm eff}$. 
First, note that the term proportional to
$f(G+\Delta)$ gives no contribution, 
because $f'(G+\Delta) = 
\frac{1}{2} (G+\Delta) (\lnbar(G+\Delta) - 1)$ vanishes at the minimum of the
potential, where $G + \Delta =0$. 
One therefore finds from eq.~(\ref{eq:Veffresummed}):
\beq
\frac{1}{v}\frac{\partial}{\partial v} 
\left (\Vresummed_{\rm eff} - V_{\rm eff} \right ) 
&=& 
-\frac{1}{16 \pi^2} 3 \lambda A(G) 
-\frac{1}{(16 \pi^2)^2} \left [
3 \lambda \Delta_1 \lnbar(G) 
+ \frac{3}{2}A(G)\,\frac{1}{v}\frac{\partial\Delta_1}{\partial v} \right ]
\nonumber \\ &&
-\frac{1}{(16 \pi^2)^3} 54 y_t^4 \left [ \frac{\lambda A(t)^2}{G}
+ y_t^2 A(t) \lnbar(t) \lnbar(G) \right ] .
\label{eq:Vmindiff}
\eeq
Note that the apparently 2-loop
term containing $A(G)$ is actually of 3-loop order, because $A(G)$
contains a factor of $G$, which at the minimum of the potential is
equal to $-\frac{1}{16\pi^2} \Delta_1 + \ldots$. Therefore, 
in eq.~(\ref{eq:Vmindiff}), to be consistent we should keep only the part 
of $\frac{1}{v} \frac{\partial\Delta_1}{\partial v}$
that involves the top Yukawa coupling:
\beq
\frac{1}{v} \frac{\partial\Delta_1}{\partial v} &=& 
-6 y_t^4 \lnbar(t)  + \ldots .
\eeq
So, one obtains from eq.~(\ref{eq:Vmindiff}):
\beq
\Delta_1 &=& \delta_1 - 3 \lambda A(G),
\\
\Delta_2 &=& \delta_2 - 3 \lambda \Delta_1 \lnbar(G) 
             + 9 y_t^4 \lnbar(t) A(G),
\\
\Delta_3 &=& \delta_3 - 54 y_t^4 \left [ \frac{\lambda A(t)^2}{G}
+ y_t^2 A(t) \lnbar(t) \lnbar(G) \right ] .
\eeq
This shows that 
the $G$-dependent
terms of eqs.~(\ref{eq:delta1})-(\ref{eq:delta3}) neatly cancel, 
up to the order that has
been calculated.

The resulting 
$\Delta_1$ does not explicitly depend on $G$, 
but depends on $m^2$ through 
$H$.
A further refinement of the minimization condition 
can be made by writing
\beq 
H &=& h + G,
\label{eq:replaceH}
\eeq
where, from eq.~(\ref{eq:defH}),
\beq
h &=& 2 \lambda v^2 
\eeq
at the minimum of the potential, and then
iteratively replacing
the $G$ dependence using
\beq
G^n \,\lnbar^p(G)
&=& 
\left (
-\frac{1}{16\pi^2}\Delta_1
-\frac{1}{(16\pi^2)^2}\Delta_2-\ldots \right)^n
\, \lnbar^p(G)
\eeq
in $\Delta_1$, $\Delta_2$, and $\Delta_3$.
(Note that logarithms of $G$ are left alone, to cancel amongst themselves.)
In doing so, one consistently drops
terms of 4-loop order as well as terms of 3-loop order that are
suppressed by $g$, $g'$ or $\lambda$. 
Thus, for example, the 1-loop contribution involving $H$ is rewritten
using 
\beq
A(H) &=&
A(h) + G A'(h) 
+ \frac{G^2}{2} A''(h) + \ldots 
\\ 
&=& 
A(h)
- \frac{1}{16\pi^2} \Delta_1 \lnbar(h) 
+ \frac{1}{(16\pi^2)^2} \left [
\frac{(\Delta_1)^2}{2h} - \Delta_2 \lnbar(h) \right ]
+ \ldots.
\eeq  
Because this is multiplied by 
$3 \lambda/16 \pi^2$ in the minimization condition,
the $\Delta_2$ term should now be dropped, but the $(\Delta_1)^2/2h$ term
is partially kept, 
because $3 \lambda (\Delta_1)^2/2h = 27 y_t^4 A(t)^2/v^2 + \ldots$. 
The self-consistent elimination of $G$ from the right side 
of the minimization condition shifts contributions 
that were originally proportional to $G^n$
up in loop order by $n$, where they can often therefore be dismissed. 
(Note that this iterative elimination of
$G$ from the right-hand side of the minimization condition 
would not have been possible without first eliminating by resummation
the terms that behave like 
$\lnbar(G)$ and $1/G$ as $G \rightarrow 0$.)

Following the procedure described above, 
I find as the condition for minimization of $\Vresummed_{\rm eff}$:
\beq
G \>=\> m^2 + \lambda v^2 &=& 
-\frac{1}{16 \pi^2} \widehat{\Delta}_1 
-\frac{1}{(16 \pi^2)^2} \widehat{\Delta}_2 
-\frac{1}{(16 \pi^2)^3} \widehat{\Delta}_3
-\ldots ,
\label{eq:resummedmincon}
\eeq
where the $\widehat{\Delta}_\ell$ depend on 
the VEV $v$ and the couplings
$\lambda, y_t, g_3, g, g'$, or on $h,t,W,Z$, 
but do not depend at all on $G$ or $m^2$. 
The results are:
\beq
\widehat{\Delta}_1 &=&  -6 y_t^2 A(t) + 3 \lambda A(h) 
+ \frac{g^2}{2} [3 A(W) + 2 W] +
\frac{g^2 + g^{\prime 2}}{4} [3 A(Z) + 2 Z],
\label{eq:Delta1hat}
\\
\widehat{\Delta}_2 &=& 
\frac{(3 g^2 - \gptwo)(33 g^4 + 22 g^2 \gptwo + \gpfour)}{
8 (g^2 + \gptwo)} I(W,W,Z) 
\nonumber \\ &&
+\left [
\frac{(9 g^4 + 66 g^2 g^{\prime 2} - 7 \gpfour) y_t^2}{6 (g^2 + \gptwo)} 
-\frac{3}{4} g^4 + \frac{1}{2} g^2 \gptwo - \frac{17}{12} \gpfour
\right ] I(t,t,Z)
\nonumber \\ &&
+ \left [ \frac{3(g^2 + \gptwo)^3}{32 (g^2 + \gptwo - 2 \lambda)}
- \frac{15}{16} (g^2 +	\gptwo)^2 + \frac{11}{4} \lambda (g^2 +  \gptwo)
- 7 \lambda^2 \right ] I(h,Z,Z)
\nonumber \\ &&
+ \left [ \frac{3g^6}{16 (g^2 - 2 \lambda)}
- \frac{15}{8} g^4 + \frac{11}{2} \lambda g^2
- 14 \lambda^2 \right ] I(h,W,W)
\nonumber \\ &&
+ y_t^2 \left [ \frac{27}{2} y_t^2 -18 \lambda \right ] I(h,t,t)
-12 \lambda y_t^2 I(0,0,t)
-15 \lambda^2 I(h,h,h)
\nonumber \\ &&
 - 6 \lambda^2 I(0,0,h) 
+ \frac{3}{4} \lambda (g^2 + \gptwo + 8 \lambda) I(0,h,Z) 
+ \frac{3}{2} \lambda (g^2 + 8 \lambda) I(0,h,W) 
\nonumber \\ &&
+ \frac{3\lambda (2 g^2 + \gptwo)^3}{2 (g^2 + \gptwo)^2} I(0,W,Z)
+ (6 y_t^4 + 3 g^2 y_t^2 - 3 g^4) I(0,t,W) 
\nonumber \\ &&
+ \left [ \frac{15}{2} \lambda g^2 - 9 g^4 
- \frac{3 \lambda g^4 (3 g^2 + 2 \gptwo)}{2 (g^2 + \gptwo)^2} 
\right ] I(0,0,W) 
\nonumber \\ &&
+ \left [\frac{9 \lambda g^4}{2(g^2 + \gptwo)} - 3 g^2 \lambda 
-\frac{21}{4} g^4 - \frac{g^2 \gptwo}{2} -\frac{103}{12} \gpfour
\right ] I(0,0,Z)
\nonumber \\ &&
+ \left [
\frac{7}{2}\lambda - \frac{g^2 + \gptwo}{2}
+\frac{3(g^2 + \gptwo)^2}{8 (2 \lambda - g^2 - \gptwo) }
\right ] A(Z)^2/v^2
\nonumber \\ &&
+ \left [
7 \lambda - \frac{35}{2} g^2 + \frac{\gptwo}{2}
+ \frac{24 g^4}{g^2 + \gptwo}
+ \frac{3 g^4}{4 (2 \lambda - g^2)}
\right ] A(W)^2/v^2
\nonumber \\ &&
+ \left [ 
\frac{6 \lambda (8 g^4 + 8 g^2 \gptwo + \gpfour)}{(g^2 + \gptwo)^2}
+ \frac{15 g^4 - 10 g^2 \gptwo - \gpfour}{g^2 + \gptwo}
\right ]
A(W) A(Z)/v^2
\nonumber \\ &&
- \frac{2 (9 g^4 - 6 g^2 \gptwo + 17 \gpfour)}{3 (g^2 + \gptwo)} A(t) A(Z)/v^2
+ 12 (y_t^2 - g^2) A(t) A(W)/v^2
\nonumber \\ &&
+\left [96 g_3^2 + 15 y_t^2 + 6 \lambda +  
\frac{9 g^4 + 90 g^2 \gptwo + 17 \gpfour}{3 (g^2 + \gptwo)}
\right ] A(t)^2/v^2
\nonumber \\ &&
+ \left [ \frac{3}{4} (g^2 + \gptwo) - 4 \lambda +
\frac{3 (g^2 + \gptwo)^2}{8 (g^2 + \gptwo - 2 \lambda)}
\right ] A(h) A(Z)/v^2 
\nonumber \\ &&
+ \left [ \frac{3}{2} g^2 - 8 \lambda +
\frac{3 g^4}{4 (g^2 - 2 \lambda)}
\right ] A(h) A(W)/v^2 \,
- 9 y_t^2 A(h) A(t)/v^2
\nonumber \\ &&
+ \biggl[
\frac{3 (g^2 + \gptwo)^3}{16 (2 \lambda - g^2 - \gptwo)}
+ \frac{\lambda (\gptwo - 23 g^2)}{2} 
- \frac{y_t^2 (63 g^4 + 30 g^2 \gptwo + 95 \gpfour)}{12 (g^2 + \gptwo)}
\nonumber \\ &&
\phantom{+}+ \frac{6 g^6}{g^2 + \gptwo} +
\frac{29}{48}g^4 + 4 g^2 \gptwo + \frac{455}{48} \gpfour 
\biggr ] A(Z)
\nonumber \\ &&
+ g^2 \biggl [
\frac{3 g^4}{8(2 \lambda - g^2)}
-\frac{21}{2} y_t^2 
- \frac{3 g^2 (\lambda + 4 g^2)}{g^2 + \gptwo}
+ \frac{605}{24} g^2 + \frac{13}{8} \gptwo 
- 8 \lambda 
\biggr ] A(W)
\nonumber \\ &&
+ y_t^2 \left [ 
16 g_3^2 - 12 y_t^2 + 24 \lambda -\frac{91}{3} g^2 
+ \frac{11}{3} \gptwo
+ \frac{64 g^4}{3 (g^2 + \gptwo)}
\right ] A(t)
\nonumber \\ &&
+ \biggl [
\frac{3 (g^2 + \gptwo)^3}{32 (2 \lambda - g^2 - \gptwo)}
+ \frac{3 g^6}{16 (2 \lambda - g^2)}
-\frac{21}{2} y_t^4 + 9 \lambda y_t^2 + 24 \lambda^2
\nonumber \\ &&
\phantom{+}- \frac{7}{4} \lambda (3 g^2 + \gptwo)
+ \frac{9}{2} g^4 + 3 g^2 \gptwo + \frac{3}{2} \gpfour 
\biggr ] A(h)
\nonumber \\ &&
+ \biggl [
24 g_3^2 y_t^4 + 9 y_t^6 + 6 y_t^4 \lambda
+ \frac{y_t^4 (18 g^4 + 87 g^2 \gptwo + 5 \gpfour)}{6 (g^2 + \gptwo)}
+ \frac{9 g^8}{32 (g^2 - 2 \lambda)}
\nonumber \\ &&
+ \frac{9 (g^2 + \gptwo)^4}{64 (g^2 + \gptwo - 2 \lambda)}
+ \frac{3 g^6 (8 g^2 + \lambda)}{8 (g^2 + \gptwo)}
- 60 \lambda^3
+ \lambda^2 (6 g^2 + 2 \gptwo - 18 y_t^2)
\nonumber \\ &&
-\frac{91}{48} y_t^2 \gpfour
+ \frac{23}{8} y_t^2 g^2 \gptwo
- \frac{27}{16} y_t^2 g^4
+ \frac{93}{16} \lambda g^4
+ \frac{7}{8} \lambda g^2 \gptwo
+ \frac{\lambda \gpfour}{16}
\nonumber \\ &&
+ \frac{199}{64} g^6 
- \frac{551}{192} g^4 \gptwo
- \frac{773}{192} g^2 \gpfour
- \frac{497}{192} g^{\prime 6}
\biggr ] v^2 ,
\\
\widehat{\Delta}_3 &=& 
g_3^4 y_t^4 v^2 \left [1036.23 - 974.20\, \lnbar(t) + 592\, \lnbar^2(t)
- 184\, \lnbar^3(t) \right ]
\nonumber \\ &&
+ g_3^2 y_t^6 v^2 \left [
-169.84 + 860.93\, \lnbar(t) - 270\, \lnbar^2(t) + 60\, \lnbar^3(t)
\right ]
\nonumber \\ &&
+ y_t^8 v^2 \biggl [
-82.91 - 753.02\, \lnbar(t) + 36\, \lnbar(h) \lnbar(t) +
\frac{657}{8}\, \lnbar^2(t) 
\nonumber \\ &&
+ 54\, \lnbar(h) \lnbar^2(t)
- \frac{225}{4}\, \lnbar^3(t)
\biggr ] + \ldots
\> .
\label{eq:Deltahat3}
\eeq
Here, the ellipses represent terms suppressed by $g$, $g'$, or $\lambda$.
The analytical versions of the decimal coefficients in eq.~(\ref{eq:Deltahat3})
are:
\beq
1036.23 &\approx& 
\frac{8170}{9} + 48 \zeta(3)
             + \frac{176}{135}\pi^4              
             + \frac{64}{9} \ln^2(2) [\pi^2 - \ln^2(2)]
             - \frac{512}{3} {\rm Li}_4(1/2) ,
\\
-974.20 &\approx& 32 \zeta(3) - 3038/3,
\\
-169.84 &\approx& 
-\frac{1172}{3}
-\frac{40}{3} \pi^2
- 96 \zeta(3) 
+ \frac{124}{15} \pi^4
+ \frac{128}{3}  \ln^2(2) [\pi^2 - \ln^2(2)]
\nonumber \\ &&
-1024\, {\rm Li}_4(1/2),
\\
860.93 &\approx& 454 + 12 \pi^2 + 240 \zeta(3),
\\
-82.91 &\approx&
\frac{3979}{8} 
+ \frac{37}{8}\pi^2 
- \frac{909}{2} \zeta(3) 
- \frac{22}{15} \pi^4
- 8 \ln^2(2) [\pi^2 - \ln^2(2)]
+ 192\, {\rm Li}_4(1/2) ,\phantom{xxxx}
\\
-753.02 &\approx&
-\frac{5145}{8} - \frac{27}{4}\pi^2 - 36 \zeta(3) .
\eeq

Having found the minimization condition in a form that does not
depend on $G$, one can now write the value of $\Vresummed_{\rm eff}$
at its minimum, again eliminating all $G$ and $m^2$ dependence by the same procedure. The result is:
\beq
\Vresummed_{\rm eff, min} &=&
\sum_{\ell = 0}^\infty \frac{1}{(16 \pi^2)^\ell} 
\Vresummed_{\ell, \rm{min}},
\label{eq:Veffmin}
\eeq
where
\beq
\Vresummed_{0, \rm{min}} &=&
\Lambda - \lambda v^4/4 ,
\\
\Vresummed_{1, \rm{min}} &=&
3 t^2 \left [\lnbar(t) - 1/2 \right ] 
- \frac{1}{2} h^2 \left [\lnbar(h) - 3/4 \right]
-\frac{3}{2} W^2 \left [\lnbar(W) + 1/6 \right]
\nonumber \\ &&
-\frac{3}{4} Z^2 \left [\lnbar(Z) + 1/6 \right] ,
\\
\Vresummed_{2, \rm{min}} &=&
-3 \lambda^2 v^2 [I(h,h,h) + I(h,0,0)]
- \frac{3}{4} \lambda A(h)^2 
\nonumber \\ &&
+ 3 y_t^2 \Bigl [ (2 t-h/2) I(h,t,t) + t I(0,0,t) + A(t)^2 \Bigr ]
\nonumber \\ &&
+ \frac{g^2 + \gptwo}{8} f_{SSV}(0,h,Z)
+ \frac{(g^2 - \gptwo)^2}{8 (g^2 + \gptwo)} f_{SSV}(0,0,Z) 
+ \frac{g^2}{4} f_{SSV}(0,h,W) 
\nonumber \\ &&
+ \frac{g^2}{4} f_{SSV}(0,0,W)
+\frac{g^2 \gptwo v^2}{4 (g^2 + \gptwo)}\left [
\gptwo f_{VVS}(W,Z,0) + g^2 f_{VVS}(0,W,0)\right ]
\nonumber \\ && 
+ \frac{g^4 v^2}{8} f_{VVS}(W,W,h)  
+ \frac{(g^2 + \gptwo)^2 v^2}{16} f_{VVS}(Z,Z,h)  
\nonumber \\ && + V_{FFV} + V_{\rm gauge} - \frac{v^2}{2} \widehat\Delta_2,
\\
\Vresummed_{3, \rm{min}} &=& 
g_3^4 t^2 \Bigl \{
184 \lnbar^3(t) 
-316 \lnbar^2(t)
+\Bigl [\frac{434}{3} -32 \zeta(3) \Bigr ] \lnbar(t)
             + \frac{293}{9} 
\nonumber \\ &&
             -64 \zeta(3)
             - \frac{176}{135}\pi^4 
             - \frac{64}{9} \ln^2(2) [\pi^2 - \ln^2(2)]
             + \frac{512}{3} {\rm Li}_4(1/2) 
\Bigr \}
\nonumber \\ &&
+g_3^2 y_t^2 t^2 \Bigl \{
-60 \lnbar^3(t) + 180 \lnbar^2(t) + 
\bigl [-94 - 12 \pi^2 - 240 \zeta(3) \bigr ] \lnbar (t)
-\frac{49}{3}
\nonumber \\ &&
+\frac{22}{3} \pi^2
- 24 \zeta(3) 
- \frac{124}{15} \pi^4
- \frac{128}{3}  \ln^2(2) [\pi^2 - \ln^2(2)]
+1024 {\rm Li}_4(1/2)
\Bigr \}
\nonumber \\ &&
+ y_t^4 t^2 
\Bigl \{
\frac{333}{4} \lnbar^3(t) 
+ \Bigl [-\frac{405}{4} -72  \lnbar(h) \Bigr ] \lnbar^2(t)
\nonumber \\ &&
+ \Bigl [\frac{5361}{8} + \frac{39}{4}\pi^2 + 36 \zeta(3) 
- 72 \lnbar(h) \Bigr ] \lnbar(t)
- \frac{3029}{16} - \frac{11}{4}\pi^2 
\nonumber \\ &&
+ \frac{22}{15} \pi^4
+ \frac{945}{2} \zeta(3) + 8 \ln^2(2) [\pi^2 - \ln^2(2)]
- 192 {\rm Li}_4(1/2)
\Bigr \},
\label{eq:V3min}
\eeq
Here $v,h,t,W,Z$ are understood to be evaluated at the solution of
the minimization condition, given by 
eqs.~(\ref{eq:resummedmincon})-(\ref{eq:Deltahat3}).
Although eqs.~(\ref{eq:Veffmin})-(\ref{eq:V3min}) have only been computed 
in Landau gauge here, the value of the effective potential at its minimum
is in principle a physical observable and does not depend on the choice of gauge fixing, unlike the
VEV itself.

Renormalization group scale invariance provides an important and non-trivial 
check on the results above. If one acts on each side 
of eq.~(\ref{eq:resummedmincon}) with
\beq
Q\frac{d}{dQ} &=& 
Q \frac{\partial}{\partial Q}  
- \gamma_\phi v \frac{\partial}{\partial v}
+ \sum_X \beta_{X} \frac{\partial}{\partial X} ,
\label{eq:ddQ}
\eeq
where $X = \{\lambda, y_t, g_3, g, g', m^2\}$,
then the results must match, up to terms of 3-loop order 
suppressed by $\lambda, g,g'$ and terms of 4-loop order.
I have checked this, using the beta functions and scalar 
field anomalous dimension $\gamma_\phi$  
given at the pertinent orders in 
eqs.~(\ref{eq:defgamma})-(\ref{eq:betalambda3})
in the Appendix. 
Equations (\ref{eq:dIxyzdx})-(\ref{eq:dIxyzdQ}) 
are also useful in conducting this check.
Similarly, I have checked that acting with eq.~(\ref{eq:ddQ})
on eq.~(\ref{eq:Veffmin}) gives 0, 
up to terms of 3-loop order 
suppressed by $\lambda, g,g'$ and terms of 4-loop order,
as required. In that check,
one has instead $X = \{\lambda, y_t, g_3, g, g', \Lambda\}$,
with the beta function for the field-independent vacuum energy density
given by eqs.~(\ref{eq:betaLambda1})-(\ref{eq:betaLambda2}).

To conclude this section, I remark on a different expansion 
procedure that eliminates the Goldstone boson dependence of the
minimization condition for the effective potential, proposed
in ref.~\cite{Patel:2011th} to avoid 
spurious gauge dependence in physical quantities
in the context of a more general gauge-fixing and finite temperature
field theory with applications to baryogenesis. 
The idea, called the ``$\hbar$ expansion" in ref.~\cite{Patel:2011th}, 
is to first write an expansion 
of the VEV in the same way as the effective potential:
\beq
v = \phi_{\rm min} = \phi_0 + \frac{1}{16\pi^2} \phi_1 + 
\frac{1}{(16\pi^2)^2} \phi_2 + 
\frac{1}{(16\pi^2)^3} \phi_3 + \ldots
\label{eq:hbarexp}
\eeq
and then to demand that, after expanding in the loop-counting parameter
(here $1/16 \pi^2$), the contributions to the
derivative of $V_{\rm eff}$ vanish separately
at each loop order. 
Then $\phi_0$ minimizes the tree-level
potential $V_0$, so that $m^2 + \lambda \phi^2_0 = 0$ and
$V_0''(\phi_0) = 2 \lambda \phi_0^2$,
with the subsequent terms in the
expansion of the VEV given by:
\beq
\phi_1 &=& -V'_1/V''_0 ,
\label{eq:phi1}
\\
\phi_2 &=& -\Bigl [V_2' + \phi_1 V_1'' + \frac{1}{2} \phi_1^2 V_0''' \Bigr ]/V_0'' ,
\label{eq:phi2}
\\
\phi_3 &=& -\Bigl [
V_3' 
+ \phi_1 V_2'' 
+ \phi_2 V_1'' 
+ \frac{1}{2} \phi_1^2 V_1''' + \phi_1 \phi_2 V_0''' +  
\frac{1}{6} \phi_1^3 V_0'''' \Bigr ]/V_0'',
\label{eq:phi3}
\eeq
etc., with all of the derivatives of $V_{\ell}$ 
on the right-hand sides evaluated
at the tree-level minimum, $\phi_0 = \sqrt{-m^2/\lambda}$. Each of the 
individual terms $V_1''$, $V_1'''$, $V_2'$, $V_2''$, and $V_3'$ diverges as 
$\phi \rightarrow \phi_0$, but one can check that the combination 
$\phi_2$ is well-behaved in this limit, and that
$\phi_3$ is well-behaved up to the approximation 
corresponding to the known calculation of $V_3$ in ref.~\cite{Martin:2013gka}. 
The result is therefore indeed free of spurious imaginary parts and
the $G \rightarrow 0$ problems, by construction. However, the result 
is organized quite differently
from that of the present paper, 
eqs.~(\ref{eq:resummedmincon})-(\ref{eq:Deltahat3}); it corresponds to a perturbative solution of the minimization conditions, expanded around the 
tree-level minimum. 
It is interesting to look at the explicit results 
for the expansion method of
ref.~\cite{Patel:2011th},
for simplicity in the limit that $y_t$ is much larger than
$\lambda$, and $g,g' = 0$:
\beq
\phi_1 &=& \frac{3 y_t^4}{2 \lambda} \left [\lnbar(t_0) -1 \right ] \phi_0
+ \ldots ,
\\
\phi_2 &=& \frac{9 y_t^8}{8 \lambda^2} \left [\lnbar(t_0) -1\right ]
\left [1 + 3\lnbar(t_0)\right ]\phi_0 + \ldots ,
\\
\phi_3 &=& \frac{27  y_t^{12}}{16 \lambda^3} 
\left [\lnbar(t_0) -1\right ]\left [-3
+ 6 \lnbar(t_0) + 5 \lnbar^2(t_0)\right ]\phi_0 + \ldots,
\eeq
with $t_0 = y_t^2 \phi_0^2/2$.
The presence of powers of $\lambda$ in the denominators is due to
$V_0'' = 2 \lambda \phi_0^2$.
This shows that, due to expanding around the tree-level VEV, 
the expansion parameter 
is effectively
$\displaystyle \frac{y_t^4}{16 \pi^2 \lambda}$,
rather than the usual perturbative expansion parameters
$\displaystyle \frac{y_t^2}{16 \pi^2}$ and 
$\displaystyle \frac{\lambda}{16 \pi^2}$.
Correspondingly, in this approach
some of the information present in the known $V_0$, $V_1$, $V_2$, and $V_3$ 
evaluated at the minimum of the full effective potential 
is postponed to the contributions $\phi_\ell$ with $\ell \geq 4$.
With the presently known approximation to $V_3$ found 
in ref.~\cite{Martin:2013gka},
the finiteness of $\phi_3$ [as the limit $\phi \rightarrow \phi_0$ is taken 
in the derivatives of $V_\ell$ in eq.~(\ref{eq:phi3})]
only works up to the order
$y_t^8\phi_0/\lambda$ 
(in an expansion of $\phi_3$ in small $\lambda/y_t^2$). 
A further calculation of subleading corrections to $V_3$ in the expansion in
$\lambda$ would be necessary
to make well-defined the $\phi_3$ contributions of order $y_t^6 \phi_0$
and $g_3^2 y_t^4 \phi_0$.
Despite these formal issues, I have checked that in practice the
numerical result of applying the expansion procedure 
of ref.~\cite{Patel:2011th} as 
in eqs.~(\ref{eq:hbarexp})-(\ref{eq:phi3}) above, with all known effective potential contributions included,
agrees very well with the expansion found in the present paper,
eqs.~(\ref{eq:resummedmincon})-(\ref{eq:Deltahat3}). For the input parameters
of eqs.~(\ref{eq:inputlambda})-(\ref{eq:inputgp}), 
the two methods agree on the predicted VEV 
to within 20 MeV. 

\section{Numerical impact\label{sec:numerical}}
\setcounter{equation}{0}
\setcounter{figure}{0}
\setcounter{table}{0}
\setcounter{footnote}{1}

\begin{figure}[t]
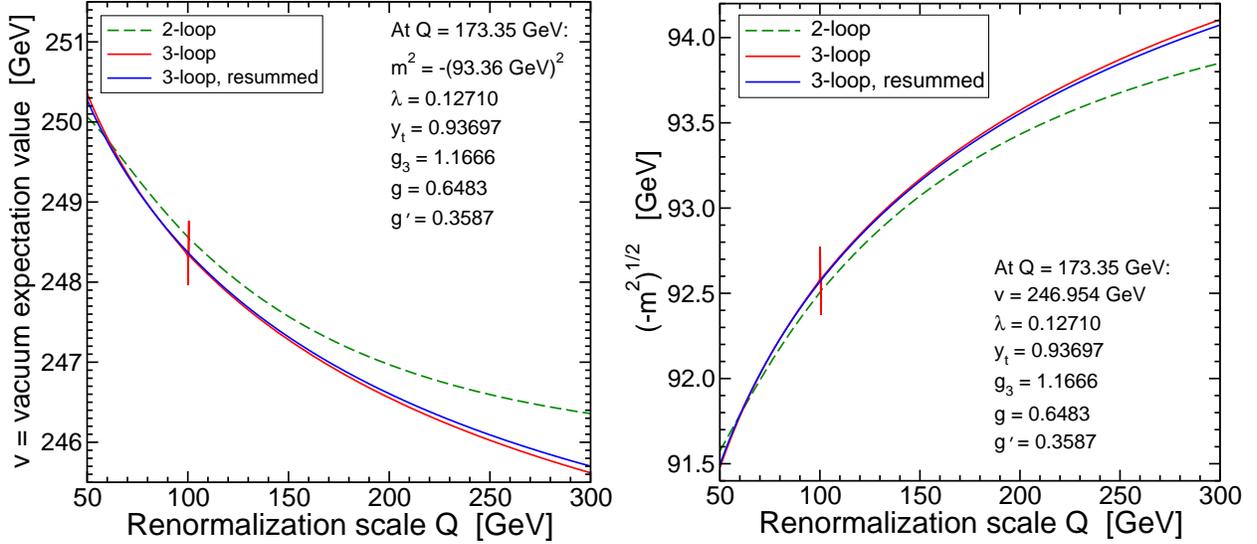

\begin{minipage}[]{0.49\linewidth}
\includegraphics[width=\linewidth,angle=0]{VEVcomp.eps}
\end{minipage}
\begin{minipage}[]{0.50\linewidth}
\begin{flushright}
\includegraphics[width=\linewidth,angle=0]{m2comp.eps}
\end{flushright}
\end{minipage}
\caption{\label{fig:num} 
Dependence of the VEV $v$ (left panel) and the Higgs Lagrangian mass parameter
$\sqrt{-m^2}$ (right panel), as a function of the renormalization scale, 
as computed from the effective potential minimization condition at 2-loop order
from ref.~\cite{Ford:1992pn}, at partial
3-loop order including also \cite{Martin:2013gka}, 
and 3-loop order after resummation using 
eqs.~(\ref{eq:resummedmincon})-(\ref{eq:Deltahat3}). 
The input parameters
are specified in eqs.~(\ref{eq:inputlambda})-(\ref{eq:inputgp}) 
at the input scale $Q=173.35$ GeV. 
In the left panel, the parameters including 
$m^2$ are run from the input scale to $Q$, and $v$ is solved for.
In the right panel, the parameters 
(including $v=246.954$ GeV at $Q=173.35$ GeV) are run to $Q$, 
and $m^2$ is solved for. The 3-loop case without resummation 
has a numerical instability associated with a failure of the iterative
solution process to converge, represented by the vertical line of
arbitrary height, for a narrow range near $Q=100.4$ GeV,
due to the $1/G$ term in the minimization condition.}
\end{figure}

The numerical effect of the resummation is very small
for almost all choices of the renormalization scale. This is illustrated in 
Figure \ref{fig:num}. In the left panel, the input parameters are specified by
eqs.~(\ref{eq:inputlambda})-(\ref{eq:inputgp})
at the input scale $Q=173.35$ GeV. These are then run using the 
full 3-loop renormalization group equations \cite{MVI}-\cite{Bednyakov:2013eba}
to the scale $Q$, where the minimization 
of the effective potential gives the VEV $v$.
Three different approximations are shown: the full 2-loop order $V_{\rm eff}$ 
of ref.~\cite{Ford:1992pn}, 
the same result including partial 3-loop order contributions
from \cite{Martin:2013gka}, and the result after resummation, using
eqs.~(\ref{eq:resummedmincon})-(\ref{eq:Deltahat3}).
(For the first two approximations, the effective potential is complex 
for all $Q\gsim 100.4$ GeV, so it is the real part that 
is minimized.) The 3-loop 
order result without resummation has a severe numerical instability, 
marked by a failure to converge of the iterative solution for $v$,
near the 
renormalization scale at which $G$ crosses through  0 (compare Figure 
\ref{fig:Grun}). 
This is 
represented by a vertical line in the figure, for a range of 
several hundred MeV in $Q$ near $Q=100.4$ GeV. (The height of this
line in the figure is arbitrary, as no iterative 
solution to the minimization condition is 
obtained.)
For other values of $Q$, the 3-loop result for $v$ is 
lower than the 2-loop result by up to a few hundred MeV, depending on 
$Q$. The resummed 3-loop result does not differ much from the 
non-resummed result, except for the numerical instability region just 
mentioned. There, the resummed result of course 
remains perfectly smooth, as it 
does not depend on $G$ at all.

The right panel of Figure \ref{fig:num} shows the result of running the same 
parameters but $v$ instead of $m^2$, 
starting again from the input scale at 173.35 
GeV, and then solving for $m^2$ at the scale $Q$ using 
eqs.~(\ref{eq:resummedmincon})-(\ref{eq:Deltahat3}). 
Again the difference between the usual
3-loop and resummed 3-loop  calculations is very small except near the scale 
$Q=100.4$ GeV where $G$ goes through 0, where the 
iterative process of solution again fails for the non-resummed case.

\section{Outlook\label{sec:outlook}}
\setcounter{equation}{0}
\setcounter{figure}{0}
\setcounter{table}{0}
\setcounter{footnote}{1}

In this paper, I have shown how issues of 
principle associated with the Goldstone
boson contributions to the effective potential 
can be resolved through resummation.
The minimization condition of 
eqs.~(\ref{eq:resummedmincon})-(\ref{eq:Deltahat3}),
and the value of the effective potential at its minimum
eqs.~(\ref{eq:Veffmin})-(\ref{eq:V3min}),
do not involve $G$ at all, and so are 
manifestly free of spurious imaginary parts
associated with $\lnbar(G)$ when $G$ is negative, 
and of divergences as $G \rightarrow 0$. 

Given the tiny numerical impact found in the previous section
for almost all renormalization scales, one 
might ask whether in practice one could not just use the usual 
prescription of minimizing the 
real part of the effective potential, taking care to choose a $Q$ to
ensure that $G$ is not too close to 0. 
Here, I note that the minimization conditions 
eqs.~(\ref{eq:resummedmincon})-(\ref{eq:Deltahat3}) 
are actually easier to 
implement in practice, because there are fewer and 
less complicated terms and 
no need to deal with imaginary parts. 
The resummation method described above is also a useful
ingredient in the analytical 2-loop calculation of the Standard Model
Higgs mass. 
Both the resummed version 
of the minimization condition 
eqs.~(\ref{eq:resummedmincon})-(\ref{eq:Deltahat3}) 
and the 2-loop Higgs mass in the Standard Model
will be implemented in a forthcoming public computer code \cite{SMDR}.

A similar resummation of Goldstone contributions can clearly be applied 
to other cases of symmetry breaking beyond the Standard Model. For 
supersymmetry, the second derivatives of $V_{\rm eff}$ have been 
used in one of the methods for approximating the lightest Higgs boson 
mass. The effective potential in minimal supersymmetry has the same
behavior \cite{Martin:2002iu,Martin:2002wn}
with respect to the tree-level squared masses of the 
Goldstone bosons at 2-loop order 
(except that $m^2_{G^\pm}$ and $m^2_{G^0}$ are slightly different from each other when not
at the minimum of the tree-level potential).
The use of second derivatives, rather than first derivatives as in 
the minimization condition, means that the numerical instabilities 
associated with choices of renormalization scale where 
$m^2_{G^\pm}\approx 0$ and $m^2_{G^0}\approx 0$ can 
be much more significant; see Figure 1 in 
ref.~\cite{Martin:2002wn} and the surrounding discussion. 
Resummation of the leading Goldstone 
contributions may also be useful in that case, although the ultimate 
resolution will come from calculating the full 
self-energy functions at non-zero external momentum invariant.

Note added: a paper \cite{Elias-Miro:2014pca} with significant overlap
appeared in the same arXiv announcement as this one.


\section*{Appendix: Effective potential results in the Standard Model}\label{sec:appendix}
\renewcommand{\theequation}{A.\arabic{equation}}
\setcounter{equation}{0}
\setcounter{footnote}{1}

The 2-loop contribution to the Landau gauge 
Standard Model effective potential was found in
ref.~\cite{Ford:1992pn}. 
The result is:
\beq
V_{2}
&=& 
V_{SSS} + V_{SS} + V_{FFS} + V_{SSV} + V_{VVS} + V_{VS} 
+ V_{FFV} + V_{\rm gauge},
\label{eq:V2decomp}
\eeq
where
\beq
V_{SSS} &=& -3 \lambda^2 v^2 [I(H,H,H) + I(H,G,G)],
\\
V_{SS} &=& \frac{3}{4} \lambda \left [A(H)^2 + 2 A(H) A(G) + 5 A(G)^2 \right ]
,
\\
V_{FFS} &=& 3 y_t^2 \Bigl [ (2 t-H/2) I(H,t,t) - \frac{G}{2} I(G,t,t) +
(t-G) I(0,G,t) 
\nonumber \\ &&
+ A(t)^2 - A(H) A(t) - 2 A(G) A(t) \Bigr ]
,
\\
V_{SSV} &=& \frac{g^2 + \gptwo}{8} f_{SSV}(G,H,Z)
+ \frac{(g^2 - \gptwo)^2}{8 (g^2 + \gptwo)} f_{SSV}(G,G,Z) 
\nonumber \\ && 
+ \frac{g^2 \gptwo}{2 (g^2 + \gptwo)} f_{SSV}(G,G,0)
+ \frac{g^2}{4} \left [ f_{SSV}(G,H,W) + f_{SSV}(G,G,W) \right]
,
\\
V_{VVS} &=& \frac{g^2 \gptwo v^2}{4 (g^2 + \gptwo)}\left [
\gptwo f_{VVS}(W,Z,G) 
+ g^2 f_{VVS}(0,W,G)\right ]
\nonumber \\ && 
+ \frac{g^4 v^2}{8} f_{VVS}(W,W,H)  
+ \frac{(g^2 + \gptwo)^2 v^2}{16} f_{VVS}(Z,Z,H)  
,
\\
V_{VS} &=& \frac{(g^2 - \gptwo)^2}{4 (g^2 + \gptwo)} f_{VS}(Z,G)
+ \frac{g^2 + \gptwo}{8} \left [ f_{VS}(Z,H) + f_{VS}(Z,G) \right ]
\nonumber \\ && 
+ \frac{g^2}{4} \left [ f_{VS}(W,H) + 3 f_{VS}(W,G) \right ]
,
\\
V_{FFV} &=& -\left [4 g_3^2 + \frac{4 g^2 \gptwo}{3(g^2 + \gptwo)}
            \right ] t f_{\overline{F}\overline{F}V}(t,t,0)
+ \frac{3 g^2}{2} \left [ f_{FFV}(0,t,W) 
+ 3 f_{FFV}(0,0,W) \right ]
\nonumber \\ && 
+ \frac{1}{24 (g^2 + \gptwo)} \biggl [
(9 g^4 - 6 g^2 \gptwo + 17 \gpfour) f_{FFV}(t,t,Z)
\nonumber \\ && 
+  8 \gptwo (3 g^2 - \gptwo) t f_{\overline{F}\overline{F}V}(t,t,Z) 
+ (63 g^4 + 6 g^2 \gptwo + 103 \gpfour) f_{FFV}(0,0,Z)
\biggr ]
,
\\
V_{\rm gauge} &=& \frac{g^2 \gptwo}{2 (g^2 + \gptwo)} f_{\rm gauge}(W,W,0)
+ \frac{g^4}{2 (g^2 + \gptwo)} f_{\rm gauge}(W,W,Z)
.
\eeq
Here the loop functions are written in terms of the 1-loop function $A(x)$ defined
in eq.~(\ref{eq:defA}) and
a 2-loop function
$I(x,y,z)$, which is invariant under interchange of any pair of $x,y,z$.
It is equal to the $\epsilon$-independent
part of the function $(16 \pi^2)^2 \hat I(x,y,z)$ defined in 
ref.~\cite{Ford:1992pn}, and it is also given  
in terms of dilogarithms in
eq.~(2.19) of \cite{Martin:2001vx}, and defined in section 2 of \cite{TSIL},
which also provides a public computer code that 
evaluates it efficiently.
The special cases of the functions $f_{SSV}$, $f_{VVS}$, $f_{VS}$, 
$f_{FFV}$, $f_{\overline{F}\overline{F}V}$, and $f_{\rm gauge}$
defined in ref.~\cite{Martin:2001vx} 
that are pertinent for the Standard Model are:
\beq
f_{SSV}(x,y,z) &=& \bigl [ -(x^2 + y^2 + z^2 - 2 x y - 2 x z - 2 y z) I(x,y,z)
+ (x-y)^2 I(0,x,y) 
\nonumber \\ &&
+ (x-y-z) A(y) A(z) + (y-x-z) A(x) A(z) 
\bigr ]/z 
\nonumber \\ &&
+ A(x) A(y) + 2 (x + y - z/3) A(z),
\\ 
f_{SSV}(x,x,0) &=& -3 A(x)^2 + 8 x A(x) - 8 x^2,
\\
f_{VVS}(x,y,z) &=& \bigl [
-(x^2 + y^2 + z^2 + 10 x y - 2 x z - 2 y z) I(x,y,z)
\nonumber \\ &&
+ (x-z)^2 I(0,x,z) + (y-z)^2 I(0,y,z) - z^2 I(0,0,z)
\nonumber \\ &&
+ y A(x) A(z) + x A(y) A(z) + (z-x-y) A(x) A(y)
\bigr ]/4 x y 
\nonumber \\ &&
+ A(x)/2 + A(y)/2 + 2 A(z) - x-y-z,
\\
f_{VVS}(0,x,y) &=& \left [(3y - 9x) I(0,x,y) - 3 y I(0,0,y)
+ 3 A(x) A(y)\right ]/4x 
\nonumber \\ && + 2 A(y) - 3 x/4 - y/2,
\\
f_{VS}(x,y) &=& 3 A(x) A(y) + 2 x A(y),
\\
f_{FFV}(x,x,0) &=& 0,
\\
f_{\overline{F}\overline{F}V}(x,x,0) &=&  4 A(x) - 8 x -6A(x)^2/x,
\\
f_{FFV}(0,x,y) &=& \left [ (x-y) (x + 2 y) I(0,x,y)
- x^2 I(0,0,x) + (x-2y) A(x) A(y) \right ]/y\phantom{xxx}
\nonumber \\ &&
+ (2y/3 - 2 x) A(y) - 2 x A(x) + x^2 - y^2, 
\\
f_{FFV}(x,x,y) &=& 2 (x-y) I(x,x,y) + 2 A(x)^2 - 4 A(x) A(y) - 4 x A(x)
\nonumber \\ &&
+ \left (2 y/3 - 4 x \right) A(y) + 4 x^2 - y^2,
\\
f_{\overline{F}\overline{F}V}(x,x,y) &=& 6 I(x,x,y) - 8 A(x) + 4 x + 2 y,
\\
f_{\rm gauge}(x,x,y) &=&
(4x-y)(12 x^2 + 20 x y + y^2) I(x,x,y)/4 x^2
\nonumber \\ && 
+  (x-y)^2 (x^2 + 10 x y + y^2) I(0,x,y)/2 x^2 y
\nonumber \\ && 
+ y (2 x^2 - y^2) I(0,0,y)/4 x^2 \,
+ x (2y-x) I(0,0,x)/2y 
\nonumber \\ && 
+ [y^2 + 18 x y -4 x^2] A(x)^2/4 x^2
+ [x^2 + 23 x y - 9 y^2] A(x) A(y)/2 x y
\nonumber \\ && 
+ [11 y + 25 x/3] A(x)
+ [11 x - 4 y/3] A(y)
+14 x^2 + 24 x y + y^2,
\\
f_{\rm gauge}(x,x,0) &=& \frac{9x}{2} I(0,0,x) - 13 A(x)^2
+ \frac{100 x}{3} A(x) - \frac{21}{2} x^2
.
\label{eq:fgauge}
\eeq
In this paper, the explicit form of $I(x,y,z)$ is not needed; instead,
the calculations rely on several identities that it satisfies. 
First, we have
\beq
I(0,x,x) &=& 2 A(x) - 2 x - A(x)^2/x,
\eeq
which has been used in writing the equations above.
Derivatives with respect to squared mass arguments are
\beq
\frac{\partial}{\partial x} I(x,y,z) &=&
\Bigl [(x-y-z) I(x,y,z) 
-2 A(y) A(z) 
+ (x-y+z) A(x) A(y)/x \, 
\nonumber \\ && 
+ (x+y-z) A(x) A(z)/x\, 
+ (y+z-x) [A(x) + A(y) + A(z)] 
\nonumber \\ && 
+ x^2 - (y+z)^2 \Bigr ]
/(x^2 + y^2 + z^2 - 2 x y - 2 x z - 2 y z),
\label{eq:dIxyzdx}
\\
\frac{\partial}{\partial x} I(0,x,x)
&=&
-A(x)^2/x^2 .
\label{eq:dI0xxdx}
\eeq
The derivative with respect to the renormalization scale $Q$ is:
\beq
Q \frac{\partial}{\partial Q} I(x,y,z) &=& 2 A(x) + 2 A(y) + 2 A(z) -2x-2y-2z.
\label{eq:dIxyzdQ}
\eeq
For making expansions in small $G$, the following 
results from eqs.~(2.29)-(2.31) of ref.~\cite{Martin:2001vx} 
are useful:
\beq
I(0,0,G) &=& G \left [-\frac{1}{2} \lnbar^2(G) + 2 \lnbar(G) 
- \frac{5}{2} - \frac{\pi^2}{6} \right ] ,
\label{eq:expI00G}
\\
I(G,G,x) &=& I(0,0,x) + 
        2 G \left [-x - I(0,0,x) + 3 A(x) - A(x) \lnbar(G) \right ]/x
+ {\cal O}(G^2),\phantom{xxxxxx}
\\
I(G,x,y) &=& I(0,x,y) + 
G \Bigl [-(x+y) I(0,x,y) - 2 A(x) A(y) 
\nonumber \\ && 
+ 3 x A(x) + 3 y A(y) 
- y A(x) - x A(y) - (x+y)^2 
\nonumber \\ && 
+ (x-y) [A(y) - A(x)] \lnbar(G) \Bigr ]/(x-y)^2
+ {\cal O}(G^2),
\\
I(G,x,x) &=& 2 A(x) - 2 x - A(x)^2/x +
G \bigl [4 + A(x)^2/(2 x^2) + 3 A(x)/x 
\nonumber \\ &&
- [1 + A(x)/x] \lnbar(G) \bigr ]
+ {\cal O}(G^2) .
\label{eq:expIGxx}
\eeq

The 3-loop contribution to the Standard Model effective potential, 
in the approximation $g_3, y_t \gg \lambda, g, g'$, 
was found in ref.~\cite{Martin:2013gka} 
(where it was written slightly differently):
\beq 
V_{3}
&=&
g_3^4 t^2 \Bigl \{
-184 \lnbar^3(t) 
+868 \lnbar^2(t)
+(32 \zeta(3) -\frac{5642}{3}) \lnbar(t)
             + \frac{16633}{9} 
\nonumber \\ &&
             + 32 \zeta(3)
             + \frac{176}{135}\pi^4 
             + \frac{64}{9} \ln^2(2) [\pi^2 - \ln^2(2)]
             - \frac{512}{3} {\rm Li}_4(1/2) 
\Bigr \}
\nonumber \\ &&
+g_3^2 y_t^2 t^2 \Bigl \{
60 \lnbar^3(t) - 360 \lnbar^2(t) + 
\bigl [814 + 12 \pi^2 + 240 \zeta(3) \bigr ] \lnbar (t)
-\frac{2393}{3}
\nonumber \\ &&
-\frac{58}{3} \pi^2
- 216 \zeta(3) 
+ \frac{124}{15} \pi^4
+ \frac{128}{3}  \ln^2(2) [\pi^2 - \ln^2(2)]
-1024 {\rm Li}_4(1/2)
\Bigr \}
\nonumber \\ &&
+ y_t^4 t^2 
\Bigl \{
-\frac{333}{4} \lnbar^3(t) 
+ [189 + 81 \lnbar(H) + 27 \lnbar(G)] \lnbar^2(t)
\nonumber \\ &&
+ [-\frac{6945}{8} - \frac{39}{4}\pi^2 - 36 \zeta(3) 
- 54 \lnbar(H) -54 \lnbar(G)] \lnbar(t)
\nonumber \\ &&
+ \frac{15767}{16} + \frac{25}{2}\pi^2 - \frac{22}{15} \pi^4
- \frac{873}{2} \zeta(3) - 8 \ln^2(2) [\pi^2 - \ln^2(2)]
\nonumber \\ &&
+ 192 {\rm Li}_4(1/2)
+ 9 \lnbar(H) + 27 \lnbar(G)
\Bigr \},
\label{eq:V3rensim}
\eeq
or, numerically,
\beq
V_{3}
&\approx&
g_3^4 t^2 \Bigl \{
-184 \lnbar^3(t) 
+868 \lnbar^2(t)
-1842.20 
\,\lnbar(t)
+ 1957.33
\Bigr \} 
\nonumber \\ &&
+ g_3^2 y_t^2 t^2 \Bigl \{
60 \lnbar^3(t) - 360 \lnbar^2(t) + 1220.93 \lnbar(t) - 780.30
\Bigr \} 
\nonumber \\ &&
+ y_t^4 t^2 \Bigl \{
-83.25 \lnbar^3(t) 
+ [189 + 81 \lnbar(H) + 27 \lnbar(G)] \lnbar^2(t)
\nonumber \\ &&
+ [-1007.63 - 54 \lnbar(H) -54 \lnbar(G)]\lnbar(t)
+ 504.51 + 9 \lnbar(H) + 27 \lnbar(G)
\Bigr \}
.\phantom{xxxxx}
\label{eq:V3rennum}
\eeq 

For the check of renormalization group invariance mentioned at the end of
section \ref{sec:min}, the scalar field anomalous dimension and
beta functions are
\beq
\gamma_\phi &=& -Q \frac{d\,\ln\phi}{dQ} \>=\> 
\sum_{\ell = 1}^\infty
\frac{1}{(16 \pi^2)^\ell} \gamma_\phi^{(\ell)},
\label{eq:defgamma}
\\
\beta_{X} &=& Q \frac{dX}{dQ} \>=\> 
\sum_{\ell = 1}^\infty \frac{1}{(16 \pi^2)^\ell} \beta_X^{(\ell)},
\eeq
with the 1-loop contributions:
\beq
\gamma_{\phi}^{(1)} &=& 3 y_t^2 - 9 g^2/4 - 3 \gptwo/4,
\\
\beta_{\lambda}^{(1)} &=&
-6 y_t^4 + 12 \lambda y_t^2 + 24 \lambda^2 - 9 \lambda g^2 - 3 \lambda \gptwo 
+ 9 g^4/8 + 3 g^2 \gptwo/4 + 3 \gpfour/8,
\\
\beta_{m^2}^{(1)} &=& m^2  [6 y_t^2 + 12 \lambda - 9 g^2/2 - 3 \gptwo/2],
\\
\beta^{(1)}_{y_t} &=& y_t \left [ 9 y_t^2/2 - 8 g_3^2 - 9 g^2/4 - 17 \gptwo/12 
\right ],
\\
\beta^{(1)}_{g_3} &=& -7 g_3^3,
\\
\beta^{(1)}_g &=& -19 g^3/6,
\\
\beta^{(1)}_{g'} &=& 41 g^{\prime 3}/6.
\eeq
The necessary 2-loop contributions are 
\cite{MVI,MVII,Jack:1984vj,MVIII}, \cite{Ford:1992pn}:
\beq
\gamma_{\phi}^{(2)} &=& 20 g_3^2 y_t^2 - 27 y_t^4/4 + 45 y_t^2 g^2/8
+ 85 y_t^2 \gptwo/24 + 6 \lambda^2 - 271 g^4/32 
\nonumber \\ &&
+ 9 g^2 \gptwo/16 + 431 \gpfour/96
,
\\
\beta_{\lambda}^{(2)} &=& -32 g_3^2 y_t^4 + 30 y_t^6 - 8 y_t^4 \gptwo/3
- 3 \lambda y_t^4 + 80 g_3^2 y_t^2 \lambda - 144 y_t^2 \lambda^2 
+ 45 y_t^2 \lambda g^2/2 
\nonumber \\ &&
+ 85 y_t^2 \lambda \gptwo/6 - 9 y_t^2 g^4/4 +
21 y_t^2 g^2 \gptwo/2 - 19 y_t^2 \gpfour/4 
-312 \lambda^3 + 108 \lambda^2 g^2 
\nonumber \\ &&
+ 36 \lambda^2 \gptwo
-73 \lambda g^4/8 + 39 \lambda g^2 \gptwo/4 + 629 \lambda \gpfour/24
+ 305 g^6/16 
\nonumber \\ &&
- 289 g^4 \gptwo/48 - 559 g^2 \gpfour/48 - 379 g^{\prime 6}/48
,
\\
\beta_{m^2}^{(2)} &=& m^2 [40 g_3^2 y_t^2 -27 y_t^4/2 
- 72 \lambda y_t^2 + 45 y_t^2 g^2/4 + 85 y_t^2 \gptwo/12  - 60 \lambda^2
\nonumber \\ &&
+ 72 \lambda g^2 
+ 24 \lambda \gptwo - 145 g^4/16 + 15 g^2 \gptwo/8 + 557 \gpfour/48],
\\
\beta^{(2)}_{y_t} &=& y_t \left [ -108 g_3^4 + 36 g_3^2 y_t^2 
- 12 y_t^4 + \ldots \right ] .
\eeq
Partial 3-loop 
contributions are needed here
only for the beta function for $\lambda$ \cite{Chetyrkin:2012rz,Chetyrkin:2013wya,Bednyakov:2013eba}:
\beq
\beta_{\lambda}^{(3)} &=& g_3^4 y_t^4 [64 \zeta(3) - 532/3] 
+ g_3^2 y_t^6 [480 \zeta(3) - 76] + y_t^8 [-72 \zeta(3) - 1599/4] + \ldots\>.
\phantom{xxxx}
\label{eq:betalambda3}
\eeq
Finally, the beta function contributions
for the field-independent vacuum energy density are:
\beq
\beta_\Lambda^{(1)} &=&  2 (m^2)^2,
\label{eq:betaLambda1}
\\
\beta_\Lambda^{(2)} &=& (12 g^2 + 4\gptwo - 12 y_t^2) (m^2)^2.
\label{eq:betaLambda2}
\eeq

{\it Acknowledgments:} I thank Hiren Patel for discussions.
This work was supported in part by the National Science Foundation 
grant number PHY-1068369.


\end{document}